\pdfoutput=1
\documentclass[aps,prl,superscriptaddress,nofootinbib,10pt,a4paper,reprint,preprintnumbers]{revtex4-2}
\usepackage{amsmath,amssymb,amsfonts,graphicx,bm,color}
\usepackage[colorlinks=true,allcolors=blue]{hyperref}
\usepackage[caption=false]{subfig}
\usepackage{tikz}
\usetikzlibrary{decorations.pathreplacing}

\usepackage{amsfonts}
\usepackage{amsbsy}
\usepackage{url}
\usepackage{enumerate}
\usepackage[justification=raggedright]{caption}
\usepackage{hhline}
\usepackage{graphicx,color}
\usepackage{amssymb,amsmath}
\usepackage{slashed}
\usepackage{hyperref}
\usepackage{braket}
\usepackage{simplewick}
\usepackage{ascmac}
\usepackage{latexsym}
\usepackage{pifont}          
\usepackage{bm}
\usepackage{comment}
\usepackage[normalem]{ulem}
\usepackage{CJKutf8}
\usepackage{cancel}
\graphicspath{{./fig/}{./}{./figs/}}
\usepackage[dvipsnames]{xcolor}
\newcommand{\red}[1]{{\color{red}#1}}

\newcommand{\zq}[1]{\textcolor{SeaGreen}{[ZQ: #1]}}

\begin{document}
\preprint{DESY-26-018}

\title{
QCD phase diagram in a magnetic field with baryon and isospin chemical potentials
}

\author{Yu~Hamada}
\affiliation{Deutsches Elektronen-Synchrotron DESY, Notkestr. 85, 22607 Hamburg, Germany}
\affiliation{Research and Education Center for Natural Sciences, Keio University, 4-1-1 Hiyoshi, Yokohama, Kanagawa 223-8521, Japan}

\author{Muneto~Nitta}
\affiliation{Department of Physics, Keio University, 4-1-1 Hiyoshi, Yokohama, Kanagawa 223-8521, Japan}
\affiliation{Research and Education Center for Natural Sciences, Keio University, 4-1-1 Hiyoshi, Yokohama, Kanagawa 223-8521, Japan}
\affiliation{International Institute for Sustainability with Knotted Chiral Meta Matter(WPI-SKCM$^2$), Hiroshima
University, 1-3-2 Kagamiyama, Higashi-Hiroshima, Hiroshima 739-8511, Japan}

\author{Zebin~Qiu}
\affiliation{Research and Education Center for Natural Sciences, Keio University, 4-1-1 Hiyoshi, Yokohama, Kanagawa 223-8521, Japan}

\begin{abstract}
Based on the chiral perturbation theory at the leading order, we present
the phase diagram of low-energy QCD 
in a magnetic field
at finite baryon and isospin chemical potentials. 
The phase diagram consists of the QCD vacuum, the chiral soliton lattice, the uniform charged pion condensation, 
an Abrikosov vortex lattice of the charged pions, 
a baryonic vortex lattice composed of neutral and charged pion vortices with their 
topological linking number being the baryon number, 
and a hybrid phase of chiral soliton and vortex lattices, with their intersections carrying the baryon number.
While the chiral soliton lattice demands ultra-strong magnetic field 
$\sim 10^{19}$ G,
the intersection phase appears at $\sim10^{17}$ G, which is more realistic in neutron stars.

\end{abstract}

\maketitle

\section{Introduction}
Understanding the phase structure of Quantum Chromodynamics (QCD) at finite density remains a fundamental challenge~\cite{Eichmann:2016yit,Schmidt:2017bjt,Fischer:2018sdj,Rothkopf:2019ipj,Guenther:2020jwe,Braguta:2023yhd,Adam:2023cee}, due to the nature of strongly correlated systems and the sign problem in lattice QCD. 
At low energies, chiral perturbation theory (ChPT)~\cite{Holt:2014hma,Hammer:2019poc,Drischler:2021kxf,Gasser:1983yg,Scherer:2002tk} provides a systematic framework using pion degrees of freedom. 
Topological effects are encoded in the conserved Goldstone-Wilczek (GW) current~\cite{Goldstone:1981kk}, which couples baryon number to gauge fields through the Wess–Zumino–Witten (WZW) term~\cite{Witten:1983tw}. The finite baryon density and electromagnetic (EM) effects are incorporated via 
$U(1)_B$ and $U(1)_{\rm EM}$ gauge fields \cite{Son:2004tq,Son:2007ny}.

 In particular, a magnetic field enriches the QCD phase diagram~\cite{Kharzeev:2015kna,Miransky:2015ava,Andersen:2014xxa,Yamamoto:2021oys,Cao:2021rwx,Iwasaki:2021nrz}, inducing phenomena such as 
magnetic catalysis (and inverse) ~\cite{Klimenko:1990rh,Bruckmann:2013oba,Aoki:2006we} and inhomogeneous chiral phases~\cite{Basar:2010zd,Nakano:2004cd}. 
In the hadronic regime, a prominent magnetic phenomenon is the neutral pion $\pi^0$ domain wall (DW) and its periodic realization, the chiral soliton lattice (CSL)~\cite{Son:2007ny,Eto:2012qd,Brauner:2016pko}. 
Through the WZW term, the CSL has lower free energy than the vacuum when the magnetic field exceeds a certain critical value.
Such an arising phase carries the baryon number as the topological charge of the GW current~\cite{Son:2007ny}, suggesting that dense hadronic matter under strong magnetic fields may be dominantly pionic~\cite{Brauner:2018mpn}. 
At higher density and stronger magnetic field, CSL transits to 
the DW Skyrmion phase in which Skyrmions emerge on top of  CSL~\cite{Eto:2023lyo,Eto:2023wul,Amari:2024fbo,Eto:2023tuu,Amari:2024mip,Copinger:2025rpo}.
Related studies include $\eta^{(\prime)}$ CSLs and non-Abelian generalizations~\cite{Huang:2017pqe,Nishimura:2020odq,Chen:2021aiq,Eto:2021gyy,Eto:2023rzd}, mixed soliton lattices of $\pi^0$ and $\eta^{(\prime)}$ mesons~\cite{Qiu:2023guy}, 
nucleation dynamics of CSLs~\cite{Eto:2022lhu,Higaki:2022gnw,Eto:2025ebz} 
as well as counterparts of CSL in QCD-like, supersymmetric, and holographic theories~\cite{Brauner:2019rjg,Brauner:2019aid,Nitta:2024xcu,Amano:2025iwi}
among others~\cite{Yamada:2021jhy,Brauner:2023ort,Canfora:2025aau}.

Realistic phenomenology also takes into account a finite isospin chemical potential, which triggers the charged pion condensation (CPC)~\cite{Ruck:1976zt,Migdal:1978az,Son:2000xc} and enables lattice simulations free of the sign problem~\cite{Kogut:2002tm,Kogut:2002zg,Kogut:2004zg,Brandt:2017oyy,Brandt:2022hwy,Abbott:2023coj}. 
Abrikosov–Nielsen–Olesen (ANO) vortices~\cite{Abrikosov:1956sx,Nielsen:1973cs} can exist in such a condensate, and they form an Abrikosov vortex lattice (AVL) in the presence of a magnetic field,
which has been studied with
either isospin or baryon chemical potential~\cite{Adhikari:2015wva,Adhikari:2018fwm,Canfora:2020uwf,Adhikari:2022cks,Gronli:2022cri,Canfora:2024mkp,Evans:2022hwr,Evans:2023hms}. 
Recently, an ANO-like vortex carrying baryon number, dubbed ``baryonic vortex'', has been discovered in the chiral limit~\cite{Qiu:2024zpg}
or with nonzero pion mass \cite{Hamada:2025inf}.
In the latter case, above a critical 
baryon chemical potential,
a local ANO vortex of charged pions $\pi^\pm$ becomes topologically linked to a global $\pi^0$ vortex, with the linking number being the baryon number~\cite{Gudnason:2020luj,Gudnason:2020qkd}.
The resulting soliton carries the baryon charge homotopic to a Skyrmion~\cite{Skyrme:1961vq,Adkins:1983ya,Adkins:1983hy,Zahed:1986qz} or a vortex-Skyrmion~\cite{Gudnason:2014hsa,Gudnason:2014jga,Gudnason:2016yix,Nitta:2015tua}, whereas realizing a model-independent description of baryons as linked vortices \cite{Eto:2024hwn}. 

\begin{figure}[tbp]
 \centering 
  \includegraphics[width=0.5\textwidth]{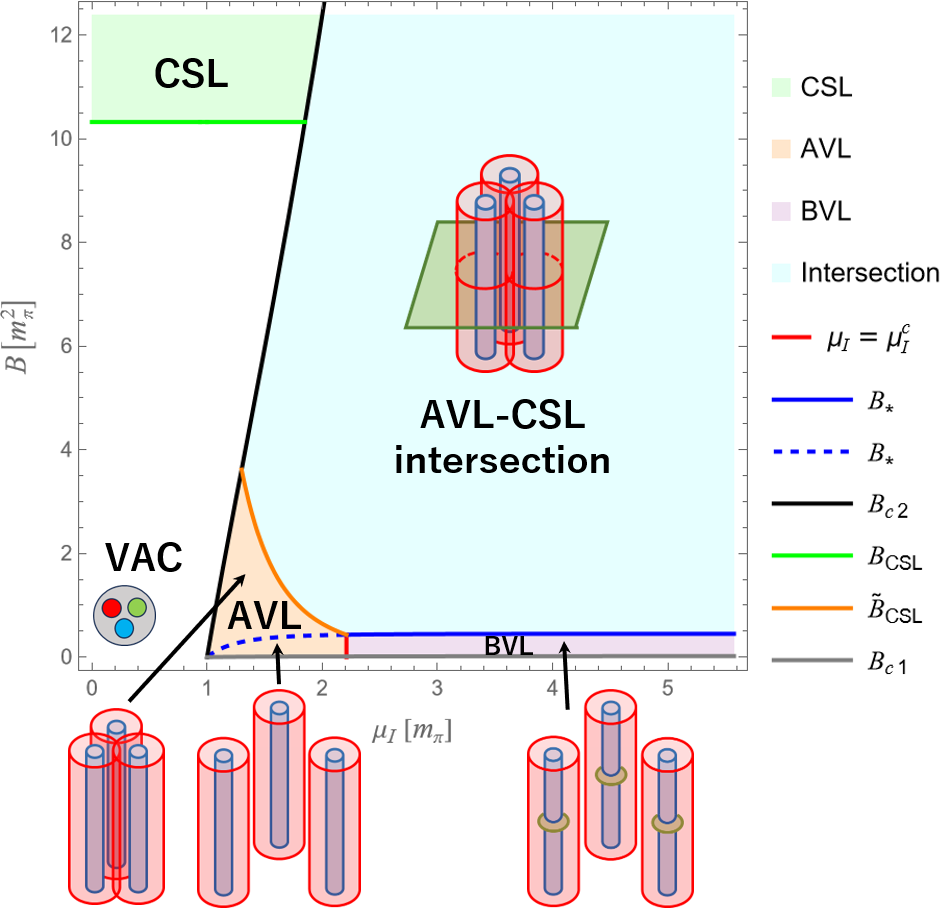}
 \caption{
 Phase diagram in one slice of $(B,\mu_B,\mu_I)$ space at the typical $\mu_B=1\text{ GeV}$. $B_\text{CSL}$ (green), $\tilde{B}_\text{CSL}$ (orange) and $\mu_I^c$ (red) depend on the value of $\mu_B$.
$B_{c1}$ (grey) almost overlaps the horizontal axis because it is far below the scale of this figure. 
The intersection of the red, blue and orange curves is located at 
$\mu_I^c = 307 \,{\rm MeV}$ 
and 
$B^c =
4.28 \times 10^{17}\text{ G } \ll B_{\rm CSL}=1.02\times 10^{19}\text{ G}$.
}
 \label{fig:phase-diagram}
\end{figure}
In this Letter, for the first time, 
we obtain
a three-dimensional phase diagram of low-energy QCD 
in terms of the magnetic field $B$, 
the baryon chemical potential $\mu_B$, 
and the isospin chemical potentials $\mu_I$,
see Fig.~\ref{fig:phase-diagram}.
Our study is performed within leading-order ChPT as a model-independent framework. Augmented with well-defined theoretical approximations, we establish all phase boundaries analytically up to one numerical constant.
In addition to established phases such as 
the QCD vacuum, the CSL, the uniform CPC, and the AVL,
our conclusive phase diagram 
includes a baryonic vortex lattice (BVL), and a CSL-AVL intersection phase.  
While the CSL requires very strong magnetic fields of order $10^{19}$ G~\cite{Son:2007ny}, the CSL–AVL intersection arises at 
$10^{17}$ G, 
thereby opening a new window for anomaly-induced baryonic phase in neutron star interiors
at lower magnetic fields than previously expected.

\section{Leading Order Chiral Perturbation}

We consider low-energy dense baryonic matter 
under a mangetic field $B$ at finite baryon chemical potential
$\mu_{B}$ 
and isospin chemical potential 
$\mu_{I}$. 
The dynamics of such a system is effectively
described by the ChPT with the WZW term.
The chiral Lagrangian for two flavors reads
\begin{equation}
\hspace{-3mm}
\mathcal{L}_{\text{chiral}}=\frac{f_{\pi}^{2}}{4}\left[\mathrm{Tr}\left(D^{\mu}\Sigma^{\dagger}D_{\mu}\Sigma\right)+m_{\pi}^{2}\mathrm{Tr}\left(\Sigma^{\dagger}+\Sigma-2\right)\right],\label{eq:Lchiral}
\end{equation}
 where $\Sigma \in SU(2)$ can be parametrized 
as
\begin{equation}
\Sigma=\exp\left(i\boldsymbol{\tau}\cdot\boldsymbol{\pi}\right)=\left(\begin{array}{cc}
\phi_{1} & -\phi_{2}^{\ast}\\
\phi_{2} & \phi_{1}^{\ast}
\end{array}\right),\;
\left|\phi_{1}\right|^{2}+\left|\phi_{2}\right|^{2}=1,
\end{equation}
with 
$\tau^{i}$ the Pauli matrices and  $\phi_1$ 
and $\phi_{2}$
complex scalar
fields 
corresponding to the 
neutral and charged pions, respectively.
We fix the pion decay constant $f_{\pi}=93\text{ MeV}$ and 
pion mass $m_{\pi}=139\text{ MeV}$.
The covariant derivative in Eq.~\eqref{eq:Lchiral} is a result of
the $U(1)$ gauging, i.e.,
\begin{equation}
D_{\mu}\Sigma=\partial_{\mu}\Sigma-
i\left(eA_{\mu}+A_{\mu}^{I}\right)\left[Q,\Sigma\right],
\end{equation}
with the charge matrix $Q=1/6+\tau^{3}/2$ and elementary charge $e$. 
It is worth clarifying
that $A_{\mu}$ represents the EM gauge field while $A_{\mu}^{I}$
is an effective description to incorporate the isospin chemical potential
$A_{\mu}^{I}=\left(\mu_{I},\,\boldsymbol{0}\right)$.
Throughout the paper, we consider static configurations with no $\partial_{0}$
and consistently, no electric field. To this end, the $\mu_{I}$ is
set to be a homogeneous input. 
In our power counting scheme,\footnote{
We adopt the same power counting scheme as Refs.~\cite{Evans:2022hwr,Evans:2023hms}, {\it i.~e.}, in terms of the momentum (denoted by $p$) expansion; $e\sim\mathcal{O}(p^1)$,  $A_{\mu}\sim\mathcal{O}(p^0)$, $\mu_I\sim\mathcal{O}(p^1)$ and $\mu_{B}\sim\mathcal{O}(p^{-1})$.
} 
the EM Lagrangian is at leading order:
\begin{equation}
\mathcal{L}_{\text{EM}}=\frac{1}{4}F^{\mu\nu}F_{\mu\nu},\quad F_{\mu\nu}=\partial_{\mu}A_{\nu}-\partial_{\nu}A_{\mu} .
\end{equation}
Specifically, we consider a uniform magnetic field along the $z$-axis $\boldsymbol{B}=B\hat{z}$.

The magnetic field brings in the effect of the chiral anomaly, which could be captured by
the gauged
WZW term: 
\begin{equation}
\mathcal{L}_{\text{WZW}}=\left(\frac{1}{2}eA_{\mu}+A_{\mu}^{B}\right)j_{B}^{\mu},
\label{eq:wzw}
\end{equation}
where $j_B^\mu$ is the topological GW current  
\begin{equation}
j_{B}=\star\frac{1}{24\pi^{2}}\mathrm{Tr}\left\{ l\wedge l\wedge l+3i e Q d\left[A\wedge\left(l-r\right)\right]\right\} ,\label{eq:jB}
\end{equation}
with $l=\Sigma^{\dagger}d\Sigma$ and $r=\Sigma d\Sigma^{\dagger}$.
The charge $N_B=\int d^3x j_B^0$ has the physical meaning of the baryon number.
In parallel to $A_{\mu}^{I}$,
the effective baryon gauge field $A_{\mu}^{B}$ encompasses the baryon chemical potential $A_{\mu}^{B}=\left(\mu_{B},\,\boldsymbol{0}\right)$. 
Once $\mu_{B}$ is introduced, the system tends to accommodate a nonzero $N_B$ couples to the $\mu_B$ through the WZW term, reducing the total energy by $-\mu_{B}N_{B}$.\footnote{
The first term in Eq.~\eqref{eq:wzw} is at subleading $\mathcal{O}(p^4)$.}
Importantly, a finite $N_{B}$ could exist only if neutral pion is involved.
There are different solitons that carry the baryon number.
One is the DW 
for which the $j_{B}$ originates from 
the second term in Eq.~\eqref{eq:jB}.
The associated phase is nothing but the CSL.
Another soliton is the Skyrmion, or technically known as the vortex Skyrmion in our context given the CPC in the bulk, distinguished from the conventional Skyrmion. 
The $j_{B}$ of such a soliton is attributed
to the $l\wedge l\wedge l$-term in Eq.~\eqref{eq:jB},
which yields a winding number of the non-trivial homotopy group $\pi_{3}(S^{3})$.

\section{Vortex and Chiral Soliton Lattices}

Let us briefly review the effect of finite $\mu_{I}$.
For $\mu_{I}>m_{\pi}$  
charged pions form the Bose-Einstein condensate, exhibiting superconductivity~\cite{Son:2000xc}. 
Characteristic scales of 
superconductors are the penetration depth $\lambda$ and the coherence length $\xi$ \cite{Adhikari:2015wva}
(see Appendix A): 
\begin{align}
\lambda^{-1} 
=ef_{\pi}\sqrt{1-\frac{m_{\pi}^{4}}{\mu_{I}^{4}}}, \quad 
\xi^{-1}
=\mu_I \sqrt{1-\frac{m_{\pi}^{4}}{\mu_{I}^{4}}},
\label{eq:lengths}
\end{align}
respectively. 
The 
Ginzburg-Landau parameter 
$\kappa \equiv \lambda/\xi 
=\mu_I/e f_\pi > 
m_\pi /e f_\pi 
\sim 4.93$ with $e^2/(4\pi)\simeq 1/137$ satisfies the condition 
$\kappa > 1/\sqrt 2$
for type-II superconductivity~\cite{Adhikari:2015wva,Gronli:2022cri}.
In parallel to metallic superconductors, there are pionic vortices
\cite{Adhikari:2015wva,Adhikari:2018fwm,Gronli:2022cri}.
Let us explain a single vortex in
the cylindrical coordinate $\left(\rho,\varphi,z\right)$.
The dependence
on the azimuthal $\varphi$ of the dynamics is entirely attributed
to the charged pion vortex, i.e., $\phi_{2}=\left|\phi_{2}\right|\exp\left(i\varphi\right)$ 
and  
$A=\rho A_{\varphi}\left(\rho,z\right)d\varphi$.
Meanwhile, $\phi_{1}$, $\phi_{1}^{\ast}$, $\left|\phi_{2}\right|$
and $A_{\varphi}$ remain functions of $\rho$ and $z$, which are to
be solved from the equation of motion (EOM) governed by $\mathcal{L}=\mathcal{L}_{\text{chiral}}+\mathcal{L}_{\text{WZW}}+\mathcal{L}_{\text{EM}}$,
with proper boundary conditions. 
Whereas details of EOM are omitted since they are identical to those in Ref.~\cite{Hamada:2025inf}, the boundary conditions deserve explication, especially on how they conserve the baryon number as a topological charge
and why it is so in search of the ground state.
There are two kinds of vortices; one is an ANO vortex
with no dependence of $\phi_1$ on $z$, which does not carry 
a baryon number
\cite{Adhikari:2015wva,Adhikari:2018fwm,Gronli:2022cri}.
The other is 
a baryonic vortex 
with $z$-dependent $\phi_1$ carrying a baryon number 
\cite{Qiu:2024zpg,Hamada:2025inf}. 
In either case,
each vortex carries a
magnetic flux quantum $\Phi_0=2\pi/e$.  

Multiple vortices repel each other and form an AVL
in an applied $B$ among the window 
$B_{c1} <B<  B_{c2}$ \cite{tinkham_introduction_2004}:\footnote{
$B_{c1}$ can be also written as 
 $ B_{c1} \sim 
\frac{\Phi_0}{4\pi \lambda^2}
\log \left(\frac{\lambda}{\xi}\right)$.
}
\begin{align}
&
B_{c1}\equiv \frac{T}{\Phi_0}
\sim \frac{e f_\pi^2}{2} 
\left(1-\frac{m_\pi^4}{ \mu_I^4}\right)
\log \left(\frac{\mu_I}{e f\pi}\right), \
 \\
&
B_{c2} \equiv 
\frac{\Phi_0}{2\pi \xi^2} 
= \frac{\mu_I^2}{e}
\left(1-\frac{m_\pi^4}{ \mu_I^4}\right).
 \label{eq:Bc12}
\end{align}
Here $T$ is the tension (energy per unit length) of a vortex and ``$\sim$'' denotes an order estimation (see Appendix B). 
The $B_{c2}$ and $B_{c1}$ 
are drawn in Fig.~\ref{fig:phase-diagram} by the black and gray curves respectively.
The AVL features the lattice spacing 
$a=\sqrt{2/\sqrt{3}n_v}$ with the 
vortex number density 
\begin{eqnarray}
 n_v = \frac{B}{\Phi_0}=\frac{eB}{2\pi}. 
 \label{eq:vortex-density}
\end{eqnarray}
At $B<B_{c1}$, vortices do not arise due to the Meissner effect, leaving a uniform CPC.  
The lattice spacing is $a\sim\xi$ 
at $B=B_{c2}$, 
and when $B>B_{c2}$, the lattice becomes so dense that 
superconductivity is broken.

The coexistence of $\mu_B$ and $\mu_I$ is essential for the physics we are to illuminate.
In the absence of CPC 
(e.~g., at $B>B_{c2}$), 
the CSL made of pure neutral pion DWs stands as the ground state for $B$ above \cite{Son:2007ny} 
\begin{equation}
B_{\text{CSL}}=\frac{16\pi f_{\pi}^{2}m_{\pi}}{e\mu_{B}},
\end{equation}
denoted by the green line in Fig.~\ref{fig:phase-diagram}.

\section{Phase Diagram in $(B,\mu_B,\mu_I)$ }

For clarity, we will present a phase diagram 
spanned by $(\mu_I,B)$ at a typical value of $\mu_B$ plus additional explanation of the dependence on varied $\mu_B$. 
The phases 
can first be divided into 
three categories 
summarized in Table \ref{tab:Bc2}.
\begin{table}[]
    \centering
    \begin{tabular}{c|cccc
    }
         Region & $B$ & $\left<\pi^\pm\right>$ & SC & vortices  
         \\\hline
         $B_{c2}<B$  & homogneous & $=0$ & no & no 
         \\
         $B_{c1}<B<B_{c2}$  & inhomogeneous & $\neq 0$&
        yes & yes 
        \\
          $B<B_{c1}$  & homogeneous & $\neq 0$&
        yes & no 
        \\
    \end{tabular}
    \caption{Categorization by the superconductivity or the charged pion condensation
    at $\mu_I>m_\pi$.}
    \label{tab:Bc2}
\end{table}
First, let us discuss the 
phases with no vortices.

{\bf $B>B_{c2}$:
QCD vacuum and CSL.}
The ground state is CSL when  $B>B_{\text{CSL}}$
is satisfied. 
Otherwise, it is the QCD vacuum.
The $B_{\text{CSL}}$
is a function of $\mu_{B}$ independent of $\mu_{I}$. Hence $B=B_{\text{CSL}}\left(\mu_{B}\right)$
turns out to be a horizontal line
in our phase diagram
in Fig.~\ref{fig:phase-diagram}.
We choose a typical $\mu_{B}= 1\text{ GeV}$ around the nucleon mass to plot the $B_{\text{CSL}}$. Such a green line
would translate downward (along the $B$-axis towards lower $B$) with an increasing $\mu_{B}$, and vice versa. 
We remark that within the region of ``QCD vacuum'' there exists the phase of traditional nuclear matter which could be described by a Skyrme crystal~\cite{Klebanov:1985qi,Chen:2021vou,Amari:2025twm}.
Such a description involves the Skyrme term, rendering the results model-dependent.
In contrast, our analysis is in the model-independent ChPT.
Moreover, within the CSL phase, there are several phase boundaries stipulating the DW Skyrmion phase~\cite{Eto:2023lyo,Eto:2023wul,Amari:2024fbo,Eto:2023tuu,Amari:2024mip,Copinger:2025rpo} and the charged pion instability~\cite{Brauner:2016pko} etc.
They are already established phenomena and not specified in the current work.

{\bf $B<B_{c1}$: Uniform CPC}. 
This region could also
accommodate the Skyrmion crystal for large  $\mu_B$, 
which is not the main focus of the current work.


Now we turn to the phase structure with superconductivity and vortices.

{\bf 
$B_{c1}<B<B_{c2}$ (Case 1): 
AVL and BVL without CSL}. 
Three scales are essential; the coherence length $\xi$, the penetration depth $\lambda$,
and the vortex lattice spacing $a$. Comparing $a$ with $\lambda$ defines a magnetic field $B_\ast$
\begin{equation}
a\sim \sqrt 2\lambda 
\leftrightarrow 
B = B_\ast \equiv  
c_b \pi ef_{\pi}^{2}\left(1-\frac{m_{\pi}^{4}}{\mu_{I}^{4}}\right) \,,
\label{eq:Bstar}
\end{equation}
with $c_b$ a constant determined later. 
$B_\ast$ separates 
\begin{eqnarray}
\begin{array}{ccc}
\mbox{
\hspace{-0.9cm}
1) Dilute 
}
a \gtrsim \sqrt{2}\lambda &\leftrightarrow&
B_{c1} \lesssim B\lesssim B_\ast ,\\
\mbox{2) Dense 
} 
\xi<a \lesssim \sqrt{2}\lambda
&\leftrightarrow&
B_\ast
\lesssim B\lesssim B_{c2} 
\end{array}
\end{eqnarray}
 through a crossover. 
In 1) 
vortices are well separated. 
In 2) the magnetic fields of vortices in contiguous cells begin to overlap each other, but vortex cores characterized by 
$\xi$ are still distanced from each other. 
We delineate $B_\ast$ with the blue curve
in the phase diagram Fig.~\ref{fig:phase-diagram}.

In addition to the above classification based on the lattice spacing, 
the nature 
of each vortex 
among a certain lattice
can be classified 
into two cases: an ANO vortex carrying no baryon number versus a baryonic vortex carrying a finite baryon number.
The difference is more clearly explained in the dilute regime, where the vortex in one cell does not interfere with others. 
Whereas the ANO vortex is
already described above 
as the charged pion condensate with a winding phase 
$\langle\phi_2\rangle e^{i\varphi}$,
the profile of a baryonic vortex
consists in the linking between charged and neutral pions, as detailed in Ref.~\cite{Hamada:2025inf} by the same authors. 
The phase associated with baryonic vortices, namely BVL, emerges on the condition that $\mu_{I}$ is above a critical value as a function of $\mu_{B}$. 
Otherwise the phase is dominated by the AVL comprised of ANO vortices: 
\begin{equation}
\text{a) AVL: }\mu_{I}<\mu_{I}^{c}\left(\mu_{B}\right)
, \quad
\text{b) BVL: }\mu_{I}>\mu_{I}^{c}\left(\mu_{B}\right).
\end{equation}
The rigorous quantification of $\mu_{I}^{c}$ is established in Ref.~\cite{Hamada:2025inf}
numerically. 
Here alternatively, we present an approximate yet analytical evaluation by imitating the BVL configuration as a simple addition of the neutral pion DW on top of the AVL. The averaged magnetic field
sustained by a single ANO vortex within an 
area bounded by the penetration
depth $\sqrt 2 \lambda$ is: 
\begin{equation}
B_{\text{vor}} = c_v \frac{\Phi_{0}}{(\sqrt2 \lambda)^{2}} 
=  c_v \pi e f_{\pi}^{2}\left(1-\frac{m_{\pi}^{4}}{\mu_{I}^{4}}\right) \label{eq:Bvor}
\end{equation}
with an order-one constant $c_v$ to be fixed. 
For the neutral pion DW to emerge inside the ANO string to form a baryonic vortex, 
such $B_{\text{vor}}$ 
surpasses the critical magnetic
field of CSL 
with the modified 
decay constant  
\begin{equation}
\tilde{B}_\text{CSL} 
\equiv \frac{16\pi\tilde f_{\pi}^2 m_{\pi}}{e\mu_{B}};\quad
\tilde{f_{\pi}}\equiv 
f_\pi \left|\phi_{1}
\right| =
f_{\pi}\frac{m_{\pi}^{2}}{\mu_{I}^{2}},
\label{eq:decay-const}
\end{equation}
which applies to regimes with a finite CPC so that 
$\left|\phi_{1}\right|\neq1$. Then the comparison yields\footnote{In this approximation, we have ignored 
the tension of a 
neutral pion vortex attached to 
the chiral soliton 
(see Appendix B).
}
\begin{equation}
B_{\text{vor}}>\tilde{B}_\text{CSL}  \leftrightarrow  
    \mu_{I}>
    \mu_{I}^{c}
    \equiv
    m_{\pi}
\sqrt[4]{
\frac{16 m_{\pi}}{c_v e^{2}\mu_{B}}+1}.
\label{eq:muIc}
\end{equation}
We find this approximation quite good when fitting with numerical results which leads to $c_v=1.06$ 
as detailed in Appendix C. 
\footnote{We use data with $\mu_I\geq3.5m_\pi$ for fitting since the numerical solutions of the baryonic vortices could not be found 
for $\mu_I \lesssim 3.5 m_\pi$ within $\mathcal{O}(p^2)$ ChPT.
However, even with that regime, we expect that $\mathcal{O}(p^4)$ ChPT with the so-called Skyrme term exhibits stable baryonic vortices and that Eq.~\eqref{eq:muIc} still holds within certain accuracy.}
We plot the $\mu_{I}^{c}\left(\mu_{B}=1\text{ GeV}\right)$ on the phase diagram in Fig.~\ref{fig:phase-diagram} with the red line. 
The transition between AVL and BVL starts inside the individual vortex, and the magnetic field 
 affects only the lattice spacing $a$ 
but not an internal structure of each vortex, as far as vortices are well separated, $B<B_\ast$.
Therefore, $\mu_{I}^{c}$
itself does not depend on $B$, yielding a vertical line.
Such a vertical line would translate leftwards (along the $\mu_{I}$-axis towards smaller
$\mu_{I}$), when $\mu_{B}$ increases, in view of Eq.~\eqref{eq:muIc}. 
In the limit $\mu_B\to \infty$, $\mu_I^c = m_\pi$ implies that the only possible vortex lattice phase is BVL. 
Conversely, for $\mu_B\rightarrow0$, the infinite $\mu_I^c \to \infty$ indicates that BVL cannot emerge. These limiting behaviors are exhibited in Appendix D.

The configuration of dense BVL around $B=B_\ast$ will be altered, 
different from a simple assemblage of well-separated baryonic vortices which were profiled in Ref.~\cite{Hamada:2025inf}.
The specific solution should be attained under carefully tailored boundary conditions, including the inter-vortex interactions, which is reserved for future work.

Combining two ways of classification, we summarize that 
phases of vortex lattices fall into four categories: 
1a) dilute AVL, 
1b) dense AVL, 
2a) dilute BVL, and
2b) dense BVL.  
We will show that 2b) does not happen.

{\bf $B_{c1}<B<B_{c2}$ (Case 2): AVL-CSL intersection}. 
A necessary condition for
a CSL to appear in phases with vortices is $B > \tilde B_{\text{CSL}}$ with noticing the modified 
$\tilde{f_\pi}$ in Eq.~(\ref{eq:decay-const}). 
This condition is valid as long as $B<B_{c2}$. 
In such a range, $\tilde{B}_\text{CSL}$ is plotted by the orange curve in Fig.~\ref{fig:phase-diagram}.
Although the CSL can be stable when the magnetic field is uniform, it is not the case if the magnetic field is 
localized by vortices:
a chiral soliton will decay in the inter-vortex region outside penetration depth $\lambda$, 
where the magnetic field is absent when vortices are well separated  $a\gg\lambda$.
Thus, the curve $B=\tilde{B}_\text{CSL}$ is invalid below $B<B_\ast$.
Such a decay of chiral solitons is avoided in denser vortex lattices with $B>B_\ast$ 
where the lattice spacing is smaller than the penetration depth 
so the magnetic fields of contiguous vortices overlap each other. 
As a result, despite the vortices, the overall magnetic field is nearly uniform, supporting the CSL. Then here is the condition for CSL to coexist with AVL and form intersections:
\begin{equation}
\max\left(\tilde B_{\text{CSL}},B_\ast\right) <B< B_{c2}.
\end{equation}
 
Let us take a closer look at what happens on the phase boundary $B_\ast$. 
In BVL, each baryonic vortex penetrates a pancake-shaped chiral soliton
whose edge is a closed 
$\pi_0$ string (in the $x$-$y$ plane).%
\footnote{
The radius of the pancakes is typically the same as $\xi\sim 1/\mu_I \sim 0.13 /f_\pi$ for $\mu_I \sim 5 m_\pi$,
so that ChPT based on the expansion with respect to $|p|\ll 4 \pi f_\pi\sim f_\pi/0.08$ is still marginally valid.
One should note that the current ChPT breaks down at $\mu_I \sim 5.6 m_\pi$ where $\rho$ meson must be taken into account.
}
The pancakes of contiguous baryonic vortices attract each other because the nearest points between two neighbored $\pi^0$ strings have opposite windings, 
and thus they sit in the same position along the vortices.
When the magnetic field $B$ is larger than the phase boundary $B_\ast$, the AVL-CSL intersection is energetically favored,
which can be regarded as a configuration after the pancakes enlarge transversely and attach each other to form an infinitely large $\pi_0$ DW.
On the other hand, $B=B_\ast$ does not prescribe a phase boundary within the AVL phase (see the dotted segment of the blue curve). 
What separates the phases of AVL and AVL-CSL intersection is $B=\tilde{B}_\text{CSL}$, at which the transition is not associated with such a deformation of chiral solitons, unlike that of BVL.

Finally, we elaborate why the orange, blue, and red curves 
should intersect at one triple point, which is achieved by $c_b = c_v$. 
When a state with vortices is adiabatically changed across the blue curve $B=B_\ast$,
only the inter-vortex distance is altered while the strength of the magnetic field inside a vortex is unchanged, so a chiral soliton can not be created or annihilated during such a process. 
Thereby, the blue curve could neither separate AVL from BVL, nor AVL from CSL-AVL intersection.
Then in 
Fig.~\ref{fig:phase-diagram}, 
the three curves, $B=\tilde{B}_\text{CSL}$, $B=B_\ast$ and $\mu_I=\mu_I^c$, must join at a common intersection
at 
\begin{align}
    (\mu_I^c,\, B^c)=
     \left(m_\pi \sqrt[4]{\frac{16 m_\pi}{c_v e^2 \mu_B} +1 },\, \frac{c_v \pi e f_\pi^2}{1  + \frac{c_v e^2 \mu_B}{16 m_\pi}}\right).
     \label{eq:triple}
\end{align}
For $\mu_B = 1$ GeV, we have 
$\mu_I^c = 307 \,{\rm MeV}$ 
and 
$B^c =
4.28 \times 10^{17} \,{\rm G}$ 
whereas 
$B_{\rm CSL} = 1.02 $ $\times 10^{19} \,{\rm G}$ 
(see Appendix D).

{\bf Baryon number densities}.

Among the phases in Fig.~\ref{fig:phase-diagram},
CSL, BVL, and AVL-CSL intersection carry baryon numbers. 
They have different profiles but the same averaged area density (in the $x$-$y$ plane) of the baryon number.
The area density of a single soliton in the CSL is 
homogeneously
$eB/2\pi$~\cite{Son:2007ny}.
In comparison, for the AVL-CSL intersection phase, it is
\begin{equation} 
\int_0^{2\pi} dz \partial_z \pi_0 \frac{e}{4\pi^2}\int d^2x F_{12} = 
 \frac{e}{2\pi} \Phi_0 \left< n_v\right> =
    \frac{e \left<B\right>}{2\pi}
    \label{eq:intersection}
\end{equation}
with the vortex number density  
$n_v$ 
defined in Eq.~(\ref{eq:vortex-density}).
The only difference from the CSL is that the magnetic field is not uniform. However, the averaged magnetic fields are equal, so is the density.
The baryon numbers of CSL and CSL-AVL intersection both come from the term $\propto d[A\wedge(l-r)]$ in Eq.~\eqref{eq:jB}.
Instead, in the BVL, 
each baryonic vortex carries $N_{\rm B}=1$ 
contributed from the $l\wedge l\wedge l$-term 
as the linking number of the ANO vortex and $\pi_0$ vortex \cite{Hamada:2025inf}.
From Eq.~(\ref{eq:vortex-density}),
the baryon number density 
within one period (along the $z$-axis) is 
\begin{equation}
    N_{\rm B} n_v = 
    \frac{e \left<B\right>}{2\pi}.
    \label{eq:baryon-BVL}
\end{equation}
The coincidence of the baryon densities in these three phases is not accidental, but reflects the topologically protected anomaly nature across distinct 
phases.

\section{Summary}

We 
presented the phase diagram in
Fig.~\ref{fig:phase-diagram}
of low-energy QCD in a magnetic field at finite baryon and isospin chemical potentials 
in leading-order ChPT. 
The baryonic vortex proposed as topologically linked neutral and charged pion vortices 
constitutes a phase called BVL, which is induced by a finite $\mu_B$ on top of an AVL, occupying the region with higher $\mu_I$ on the phase diagram. Also revealed is a phase of AVL-CSL intersection which borders vacuum, CSL, AVL, and BVL, dominating the large $B$ regime.
Within a legitimate approximation, all phase boundaries are obtained 
analytically, except for one numerical constant. 
The baryon number densities 
of CSL, BVL, and AVL-CSL were shown to coincide, reflecting their common anomaly-induced topological origin.

The characteristic magnetic field in the current study typified by $B^c\sim10^{17} \text{G}$
applies to the context of heavy-ion collisions. However, the $\mu_I^c \sim 300$ MeV is rather large compared to the typical $\mu_I \lesssim 10~\mathrm{MeV}$.
Nonetheless, neutron stars typically feature
$\mu_I \sim 50$--$150~\mathrm{MeV}$, with the possibility to reach 
$\mu_I \sim 200$--$300~\mathrm{MeV}$ in the cores. 
In the same context, $\mu_B\sim 1$--$2~\mathrm{GeV}$
and strong magnetic fields 
$B \sim 10^{16}$--$10^{18}~\mathrm{G}$ 
can exist. 
These facts suggest the prominent relevance of
the AVL-CSL intersection 
and BVL 
in the core of neutron stars.
In such a regime, various exotic phases come into play, e.g., pion condensed, hyperonic, and Fulde–Ferrell–Larkin–Ovchinnikov phases \cite{Son:2000xc}. 
Understanding their competition in the phase diagram remains a future task.

{\bf Acknowledgments.}
We thank Naoki Yamamoto and 
Geraint W. Evans for useful comments.
This work is supported in part by JSPS Grant-in-Aid for Scientific Research KAKENHI Grant No. JP22H01221 and JP23K22492 (M.~N. and Z.~Q.)
and by the Deutsche Forschungsgemeinschaft under Germany's Excellence Strategy - EXC 2121 Quantum Universe - 390833306.
The work of M.~N. is supported in part by the WPI program ``Sustainability with Knotted Chiral Meta Matter (WPI-SKCM$^2$)'' at Hiroshima University.

\bibliographystyle{jhep}
\bibliography{references}

\newpage
\onecolumngrid
\begin{center}
{\bf End Matter}
\end{center}
\twocolumngrid

\begin{appendix}

\section{Appendix A: Characteristic Lengths}\label{sec:Higgs}

We derive the effective theory for the Higgs mode and obtain the coherence length.
For simplicity, we turn down the gauge field and reduce the Hamiltonian to
\begin{align}
&\mathcal{H}_{\text{chiral}}=
\frac{f_{\pi}^{2}}{2}  \bigg\{
|\nabla\phi_{1}|^2 + 
|\nabla\phi_{2}|^2
 -\mu_{I}^{2}|\phi_2|^{2}
 \nonumber \\
 & \quad 
 +m_{\pi}^{2}\left(2-\phi_{1}-\phi_{1}^{\ast}\right)\bigg\}.
 \label{eq:Higgs}
\end{align}
The coherence length $\xi$ is proportional to
the inverse of the effective Higgs mass, which is inferred from the expansion of $\phi_2$ around its vacuum expectation value:
\begin{equation}
|\phi_2| =  \sqrt{1 - m_\pi^4/\mu_I^4} + \delta,\;
\phi_1 = \sqrt{ 1- \left( \sqrt{1 - m_\pi^4/\mu_I^4} + \delta \right)^2}.
\end{equation}
We expand Eq.~\eqref{eq:Higgs} up to $\mathcal{O}(\delta^2)$, attaining the effective Hamiltonian density 
for the Higgs mode as
\begin{eqnarray}
\mathcal{H}_\delta
= \frac{f_\pi^2\mu_I^4}{2m_\pi^4} 
\left[ 
(\nabla \delta )^2
+ \xi^{-2}\delta ^2 
\right]    ,
\end{eqnarray}
with $\xi$ the coherence length:\footnote{
This is the same as Ref.~\cite{Adhikari:2015wva}.
It differs from that in Ref.~\cite{Gronli:2022cri}
but the asymptotic behavior at large $\mu_I$ 
is the same: $\xi^{-2} \sim \mu_I^2$.
}
\begin{eqnarray}
   \xi^{-2}  
   = \mu_I^2 \left(1 -\frac{m_\pi^4}{\mu_I^4} \right) .
\end{eqnarray}

On the other hand,
the penetration depth $\lambda$ is
proportional to the inverse  
mass of $A_{\mu}$, 
which can
be read off from the  
mass term in the form of $m_{A}^{2}A_{\mu}A^{\mu}/2$
among Eq.~\eqref{eq:Lchiral} after expansion near the 
uniform ground state:
\begin{align}
\lambda^{-2} = m_{A}^2
= e^2f_{\pi}^2\left|\phi_{2}
\right|^2
=e^2f_{\pi}^2\left(1-\frac{m_{\pi}^{4}}{\mu_{I}^{4}}\right).
\end{align}

\section{Appendix B: Vortex tension}

\begin{figure}[h]
    \centering \includegraphics[width=0.9\linewidth]{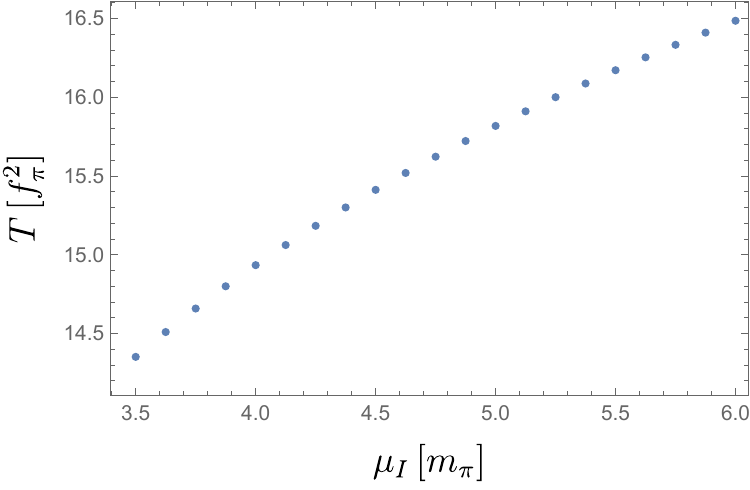}
    \caption{
    ANO vortex tension by numerical calculations.
    }
    \label{fig:tension}
\end{figure}

We evaluate 
the string tensions {\it i.e.}, the energies per unit length, of an ANO (charged pion) vortex and 
a neutral pion vortex.
 The former can be calculated as
\begin{align}
T 
& \simeq \int d^2x \left[\frac{f_\pi^2}{2}\left(\left|\nabla\big[|\phi_2|\exp{\left(i\varphi\right)}\big]\right|^2  
+ \left|\nabla |\phi_1|\right|^2\right)
+ \frac{1}{2} B^2\right]
\nonumber\\
& \simeq \frac{f_\pi^2}{2}\left(1-\frac{m_\pi^4}{ \mu_I^4}\right) \int_0^{2\pi} d\varphi \int_{\xi}^{\lambda}  \frac{d\rho}{\rho}+c\,,\nonumber \\
  & = {\pi f_\pi^2} \left(1-\frac{m_\pi^4}{ \mu_I^4}\right) 
  \log \left(\frac{\mu_I}{e f_\pi}\right)
  +c,
  \label{eq:c}
\end{align}
where we have used the facts 
that the vortex behaves 
as a global vortex 
inside the penetration depth 
$\rho \lesssim \lambda $ 
and $|\phi_2| \sim  1-\frac{m_\pi^4}{ \mu_I^4}$ at $\rho \gtrsim \xi$.
Here, $c$ includes several contributions:  
the gradient energies of 
 $\phi_{1,2}$ inside the coherence length 
 $\rho  \lesssim \xi$, 
 an energy of the magnetic field, and the negligible (exponentially suppressed) contributions outside 
 $\rho \gtrsim \lambda$. 
For $\mu_I/m_\pi=4.0$, the first term in Eq.~\eqref{eq:c} is analytically computed $\sim 10 f_\pi^2$, compared to the numerically obtained $T\sim 15 f_\pi^2$ (see Fig.~\ref{fig:tension}), 
indicating that 
Eq.~\eqref{eq:c} gives a good order estimation with $c\sim\mathcal{O}(1)$. 
The $c$ in the evaluation of $B_{c1}$ has been ignored in Fig.~\ref{fig:phase-diagram}
because $B_{c1}$ is much smaller than other scales regardless of $c$.

The string tension  of a 
neutral pion vortex 
attached to the boundary of a chiral soliton can 
be estimated as
\begin{align}
T_{\pi_0} 
& \simeq \int d^2x \left[\frac{f_\pi^2}{2}\left(\left|\nabla\big[|\phi_1|\exp{\left(i\varphi\right)}\big]\right|^2  
+ \left|\nabla |\phi_2|\right|^2\right)
\right]
\nonumber\\
& \simeq \frac{f_\pi^2}{2}\frac{m_\pi^4}{ \mu_I^4}\int_0^{2\pi} d\varphi \int_{\xi}^{m_\pi^{-1}}  \frac{d\rho}{\rho} 
+ b\,,\nonumber \\
  & = 
  \frac{\pi \tilde f_\pi^2}{2} 
  \log \left[\frac{\mu_I^2}{m_\pi^2} 
  \left(1-\frac{m_\pi^4}{\mu_I^4}\right)\right]
  +b, \label{eq:pi0-tension}
\end{align}
where the vortex exists 
within the core size $m_\pi^{-1}$ of the chiral soliton, and it becomes vacuum outside. 
$b$ is a contribution 
inside $\xi$. 
$T_{\pi_0}$ is parametrically 
small compared with the energy of a chiral soliton and therefore ignored in deriving Eq.~\eqref{eq:muIc}.

\section{Appendix C: Fit for $c_v$}

\begin{figure}[tbp]
    \centering \includegraphics[width=0.9\linewidth]{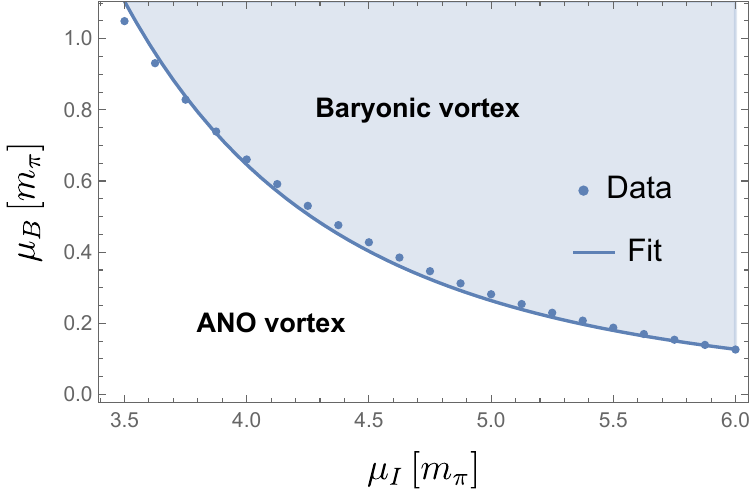}
    \caption{
    The fitting 
    for $c_v$.
 Baryonic vortices constitute the BVL while ANO vortices comprise the AVL.}
    \label{fig:fit}
\end{figure}

We fit the numerical result 
of the phase boundary between AVL and BVL established in Ref.~\cite{Hamada:2025inf} 
with the function $\mu_I =\mu_I^c(\mu_B)$ given in Eq.~\eqref{eq:muIc}, 
leading to $c_v=1.06$.
The region with larger $\mu_I/m_\pi$ gives a better fit,
because
the tension of a neutral pion vortex in  
Eq.~(\ref{eq:pi0-tension}) 
is more negligible for larger $\mu_I$.

Note that this numerical study is done with $\mu_I \gtrsim 3.5 m_\pi$.
For $\mu_I \lesssim 3.5 m_\pi$,
we found no stable solution within our numerical approach as stated in Ref.~\cite{Hamada:2025inf}. However, that does not rule out the possibility of the existence of solutions. 
Rigorously speaking, we do not know what happens to the baryonic vortex for $\mu_I \lesssim 3.5 m_\pi$ within $\mathcal{O}(p^2)$ ChPT.

\section{Appendix D: Further details on the phase diagram}
We present several scales of $\mu_{I}$ and $B$ given by intersections of curves in the phase diagram Fig.~\ref{fig:phase-diagram}. We adopt
$m_\pi =139 \,{\rm MeV}$,
$f_\pi = 93 \,{\rm MeV}$,
$e \sim 0.303$. 
As a reference, we take
$\mu_B = 1$ GeV.
For the unit conversion of the magnetic field, $1~\mathrm{MeV}^2 \simeq 5.13 \times 10^{13}~\mathrm{G}$.

First, the intersection of
 $\tilde B_{\rm CSL}$ and $B_\ast$ reads 
 Eq.~(\ref{eq:triple}).
The specific value is $(307\text{ MeV},4.28\times10^{17}\text{ G})$.
This is a triple point at which the line 
$\mu_I=\mu_I^c$ also ends.

Next, we show the intersections 
between $B_{c2}$ and 
the other phase boundaries:
\begin{eqnarray}
1) &&    B_{c2} = B_{\rm CSL}  \nonumber\\
&&    \leftrightarrow
 \mu_I^2 
 = \frac{8\pi m_\pi f_\pi^2}{\mu_B} +
 \sqrt{\frac{64 \pi^2 m_\pi^2 f_\pi^4}{\mu_B^2} 
 + m_\pi^4}. \quad\quad
\end{eqnarray}
The intersection is $(257\text{ MeV},1.02\times10^{19}\text{ G})$.

\begin{eqnarray}
2) &&    B_{c2} = \tilde B_{\rm CSL} 
\nonumber\\    \leftrightarrow 
&& \mu_I^2 
 = \sqrt[3]{\frac{8\pi m_\pi^5 f_\pi^2}{\mu_B}
 + \sqrt\Delta}
 + \sqrt[3]{\frac{8\pi m_\pi^5 f_\pi^2}{\mu_B}
 - \sqrt \Delta } \nonumber \\
&& \Delta \equiv\frac{64\pi^2 m_\pi^{10}f_\pi^4}{\mu_B^2} -\frac{m_\pi^{12}}{27}.
\end{eqnarray}
The intersection is $(181\text{ MeV},3.59\times10^{18}\text{ G})$.

\begin{eqnarray}
 3) &&   B_{c2} = B_\ast(\mu_I\rightarrow\infty)= c_b \pi ef_\pi^2 \nonumber\\
 &&   \leftrightarrow
    \mu_I^2 
    = \frac{1}{2} 
    \left(c_b \pi e^2 f_\pi^2+ \sqrt{c_b^2 \pi^2e^2f_\pi^4 
    + 4 m_\pi^4}\right).
\end{eqnarray}
The virtual intersection is 
$(145 \text{ MeV},4.46 \times 10^{17} \text{ G})$.
Typical scale of BVL magnetic field can be estimated by such asymptotic $B_\ast(\mu_I\rightarrow\infty)=c_b \pi e f_\pi^2$. The extension in the backward direction (towards smaller $\mu_I$) of such a horizontal line $B=B_\ast(\mu_I\rightarrow\infty)$ has the (virtual) intersection with the $B_{c2}$ shown above.

Also, $B_\ast$ intersects with $B_{c1}$ at
\begin{align}
    B_{c1} = B_\ast \leftrightarrow 
    \mu_I &= ef_\pi \exp (4\pi c_b) 
    \sim 17 \, {\rm TeV} ,
\end{align}
which is quite large and out of the range of ChPT.

\begin{figure}[h]
\centering
\scalebox{0.46}{{\includegraphics{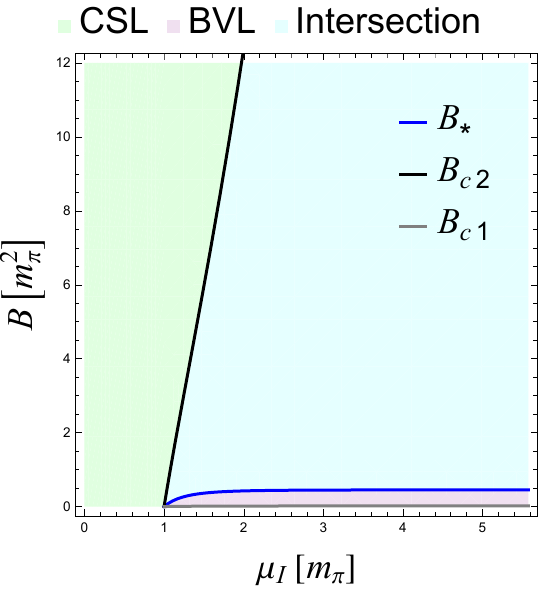}}}
\scalebox{0.46}{{\includegraphics{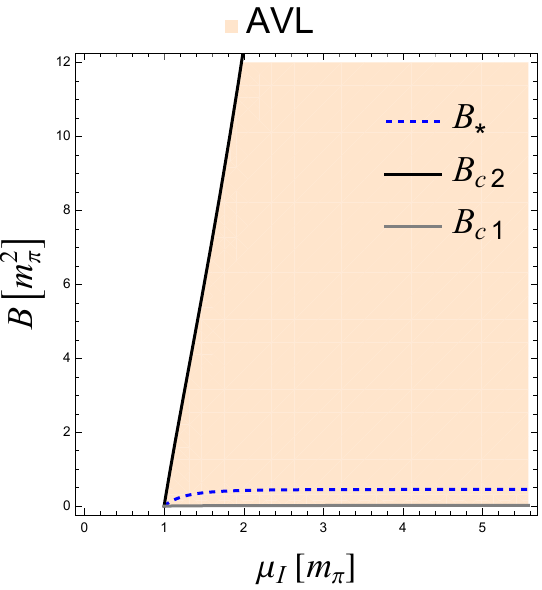}}}
\caption{Phase diagrams at $\mu_B\rightarrow\infty$ (left) and $\mu_B=0$ (right).
} 
\label{fig:phase_extreme}
\end{figure}

Eventually, we discuss phase diagrams at extreme values of $\mu_B$, as
shown in 
Fig.~\ref{fig:phase_extreme}.
In the limit $\mu_B\to \infty$, the fact $\mu_I^c \rightarrow m_\pi$ implies that the only possible vortex lattice for $B<B_\ast$ is the BVL. The transition between the BVL and the CSL-AVL intersection becomes essentially controlled by the lattice density.  
Meanwhile, the phase without vortices is dominated by the CSL when $B>B_{c2}$ and by the uniform CPC when $B<B_{c1}$.
All ground states except CPC carry a baryon number.
On the contrary, for $\mu_B\rightarrow0$, the infinite $\mu_I^c \to \infty$ means that BVL cannot emerge. Actually, there is no involvement of $\pi^0$, so the vortex phase is purely the AVL that satisfies $B_{c1}<B<B_{c2}$ with $B=B_\ast$ prescribing only a crossover (not a phase transition). The region with $B<B_{c1}$ remains the uniform CPC. No state carries a baryon number in this limit.

\end{appendix}
\end{document}